\documentclass[a4paper,11pt]{article}
\usepackage{jinstpub} 
\usepackage{lineno}
\usepackage{siunitx}
\DeclareSIUnit \photoelectron {PE}
\DeclareSIUnit \photon {photon}
\DeclareSIUnit \bar {bar}
\DeclareSIUnit \event {event}

\title{\boldmath Measurement of the Absolute Photon Detection Efficiency of the DUNE Far Detector Vertical Drift X--ARAPUCAs}

\collaboration[c]{on behalf of DUNE collaboration}

\author{S. Manthey Corchado}
\affiliation{CIEMAT, \href{https://neutrinos.portales.ciemat.es/}{neutrino group}\\
Av. Complutense 40, Madrid, Spain}

\emailAdd{sergio.mathey@ciemat.es}

\abstract{The Deep Underground Neutrino Experiment (\href{https://www.dunescience.org/}{DUNE}) is a long--baseline (\SI{1300}{\kilo\metre}) neutrino experiment hosted at the Fermi National Accelerator Laboratory (\href{https://www.fnal.gov/}{FNAL}). It aims to measure neutrino mass ordering and CP violation through neutrino oscillations from a characterized muon neutrino beam. DUNE will deploy four Liquid--Argon Time--Projection--Chamber (LArTPC) detectors with a total mass of approximately \SI{70}{\kilo\tonne}. The reconstruction of particle interactions, both from the beam and external neutrino sources is achieved by collecting two distinct interaction signals: ionization electrons with the Time Projection Chamber (TPC) and scintillation photons (\SI{127}{\nano\metre}) with the Photon Detection System (PDS). Regarding the latter, to fulfil the physics requirements of the experiment, a uniform and efficient collection of the argon scintillation light across the \qtyproduct{62 x 15 x 14}{\meter} detector volume is required to achieve an average detected light yield of at least \SI{20}{\photoelectron\per\mega\eV}. For the case of DUNE's far detector module with vertical drift direction (FD--VD), the system relies on 672 X--ARAPUCA (XA) tiles, which trap photons inside their highly reflective box by shifting VUV light to visible wavelengths. An intensive R\&D campaign, involving multiple international institutions, has optimized the design and component selection for the next--generation PDS modules, which have been tested in liquid argon using a dedicated cryogenic setup developed at CIEMAT to evaluate their photon detection efficiency~(PDE). Several configurations have been chosen to evaluate the possible design improvements in terms of different light--trapping strategies and reflectiveness.}


\arxivnumber{} 

\begin{document}
\maketitle
\flushbottom

\section{The DUNE Far Detector Vertical Drift Photon Detection System}\label{sec:introduction}

\begin{figure}[ht!]
    \centering
    \includegraphics[width=0.47\linewidth]{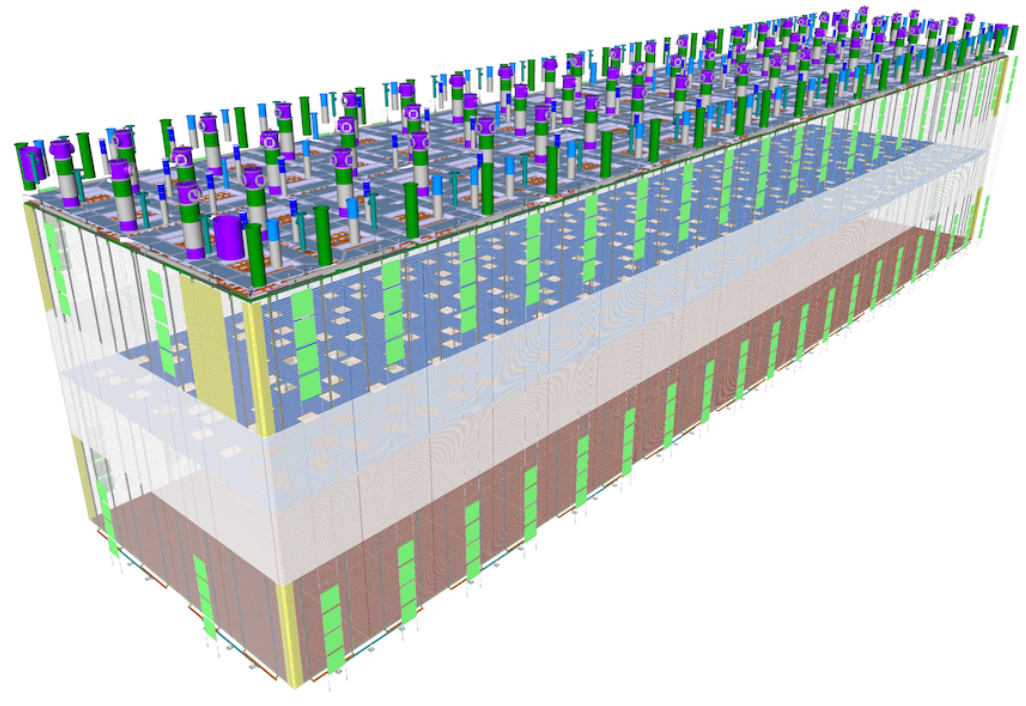}
    \caption{DUNE 17 kt FD--VD module, showing the cathode plane (blue) in the middle of the detector and two anode planes (bottom and top of the detector), as well as the different XA tiles mounted on the cathode (grey) and membrane walls (green).}
    \label{fig:DUNE-FD1}
\end{figure}

The Deep Underground Neutrino Experiment (DUNE)~\cite{TDR1_1807.10334} is a next--generation dual--site experiment that aims to provide a precise measurement of neutrino oscillation parameters~\cite{TDR2_2002.03005}. DUNE's far detector~(FD) complex consists of four \SI{17}{\kilo\tonne} liquid argon time projection chamber~(LArTPC) modules which make it possible to reconstruct 3D neutrino interactions with high precision. The first FD module to be installed is designed to implement the vertical drift~(VD) technology, in which ionization charges drift vertically in the presence of an electric field towards the anode planes positioned parallel to the Earth's surface. This module implements a central cathode plane (hosting 320 of the PDS modules) and two anode planes generating two drift volumes of approximately \SI{7}{\meter} drift distance (see Figure~\ref{fig:DUNE-FD1}). The remaining 352~PDS modules are placed outside of the transparent regions of the field cage in the so--called membrane walls of the cryostat. Liquid argon~(LAr) is an ideal target for neutrino detection because it produces abundant VUV scintillation light\footnote{LAr emits \SI{51000}{\photon\per\mega\eV} when excited in the absence of an electric drift field.}~\cite{Doke_2002}. The energy deposited by a particle interaction generates singlet and triplet states of Ar excited dimers ($\text{Ar}_{2}^{\ast}$) that de--excite with characteristic times of \SI{7.1(1.0)}{\nano\second} and about \SI{1.66(10)}{\micro\second}~\cite{Hitachi1983}, respectively. In this process, VUV photons (\SI{127}{\nano\metre}) are emitted which are collected to enhance DUNE's physics capabilities by providing accurate timing information and improved energy reconstruction of the interactions, as well as trigger capabilities for non--beam events. To fulfil DUNE's supernovae physics neutrino program (and other studies such as nucleon decays), DUNE requires a minimum average detected light yield of \SI{20}{\photoelectron\per\mega\electronvolt}~\cite{dunecollaboration2023dune}. The PDS of DUNE FD--VD consists of light collector modules, called X--ARAPUCA~(XA)~\cite{machado2018x}. The FD--VD module will use two different XA variants: a double--sided configuration, to equip the cathode plane (which collects scintillation photons from both TPC volumes), and a single--sided variant to equip the membrane planes. Mounting these two sets of XAs, instead of just equipping the cathode plane, ensures a more homogeneous response across the full drift volume. To validate DUNE's technology both XA's for the VD are required to have a minimum Photon Detection Efficiency (PDE) of 2\% (with a target value of 3\%)~\cite{dunecollaboration2023dune}.

\begin{figure}[ht!]
	\centering
	\includegraphics[width=.44\textwidth]{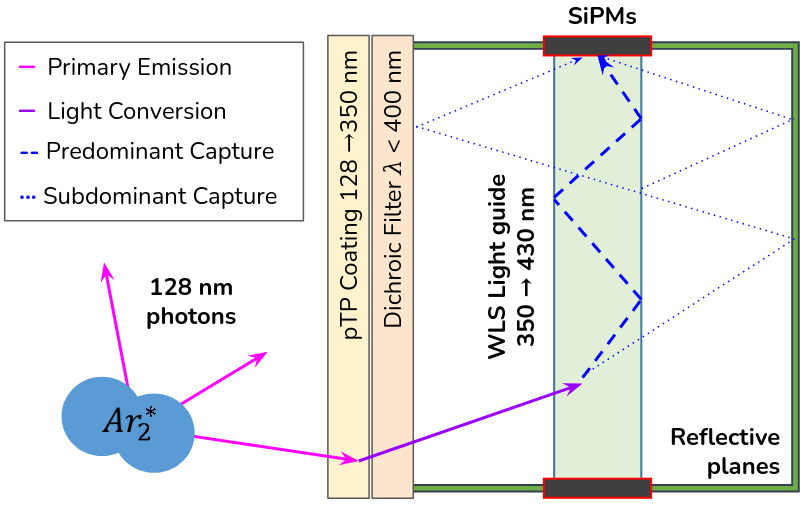}
    \includegraphics[width=.54\textwidth]{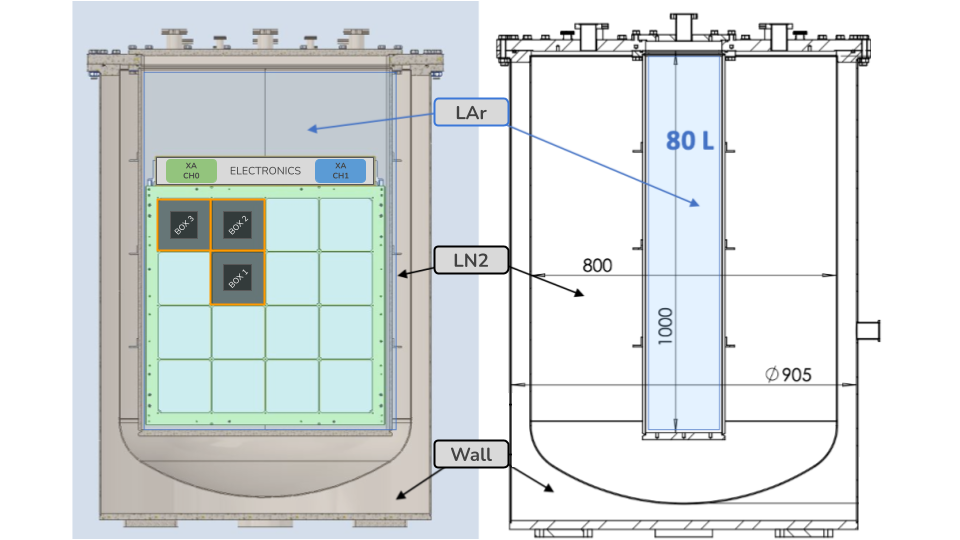}
    \caption{(left) Schematic of the XA working principle. (right) Representation of the VD--XA module inside CIEMAT's Cryogenic setup.\label{fig:xa_graphic}}
\end{figure}

The XA, as a light--sensor, follows the concept of photon--trapping inside a highly reflective box (see Figure~\ref{fig:xa_graphic} (left)). The box entrance is covered by a transparent substrate with  p–Terphenyl (pTP) coating on the outer side converting the incident VUV photons to the visible regime ($\sim$\SI{350}{\nano\meter}) and a dichroic--filter coating on the inner side preventing wavelength-shifted photons from escaping. Inside the box there is a wavelength--shifting bar~(WLS bar) that further shifts the photons ($\sim$\SI{430}{\nano\meter}) to surpass the transmittance cut-off value (\SI{400}{\nano\meter}) of the dichroic filter and guides them by total internal reflection (critical angle $\theta\sim56^{\circ}$) to SiPM photosensor arrays constituted by boards of 6 SiPMs, where each sensor has an active surface of $\sim\qtyproduct{6 x 6}{\milli\metre}$. In the specific case of the VD--XA design ($\sim\qtyproduct{60 x 60}{\centi\metre}$), 160 SiPMs are mounted on eight flex--boards and positioned around the WLS bar. These are ganged in groups of five and deliver output data across two channels connected to the readout cold--amplifier; the signals are routed to the warm electronics by signal leading boards custom--designed by \href{https://www.mi.infn.it/it/}{INFN} Sezione di Milano~\cite{gallice2021development}. The XA frame consists of a set of brackets that host a spring--loaded mechanism enhancing the contact between the SiPM's surfaces and the WLS bar. The structure also holds a grid supporting sixteen pTP--coated substrate--tiles that convert the incident photons.

\begin{table}[ht!]
	\centering
    \resizebox{.58\textwidth}{!}
    {
    \begin{tabular}{llccr}
        \hline
        & Configuration & Dichroic Filter & WLS bar & Type \\ \hline \hline
        1 & DF-XA & Yes &  A & Single--Sided \\
        2 & DF-XA-DS & Yes & A & Double--Sided \\
        3 & noDF-XA & No  & A & Single--Sided \\
        4 & noDF-XA-DS & No  & A  & Double--Sided \\ 
        5 & noDF-XA\_24mg & No & B & Single--Sided \\\hline
    \end{tabular}
    }
    \resizebox{.40\textwidth}{!}
    {
    \begin{tabular}{ccc}
        \hline
        WLS bar & A & B \\ \hline \hline
        Length~(\unit{\milli\metre})    & 605 & 607 \\
        Width~(\unit{\milli\metre})     & 605 & 605 \\
        Thickness~(\unit{\milli\metre}) & 3.8 & 5.8 \\
        Chromophore~(\unit{\milli\gram\per\kilo\gram}) & 80  & 24  \\\hline
    \end{tabular}
	}
	\caption{(left) Tested XA configurations. (right) WLS bar dimensions and chromophore concentrations were considered for the ProtoDUNE runs. \label{tab:wls_configurations}}
	
\end{table}

Despite the ideal containment strategy of the photons explained in the XA's working principle, recent simulations of the VD's XA design predict a worsening in PDE with the implemented ZAOT filters due to a non--ideal transmittance behaviour which differs from the theorised sharp cut--off profile. The response of the XA with non--coated substrates has shown an improvement with respect to the original coated design. Additionally, the impact of the WLS bar's thickness and chromophore concentration on the XA's efficiency has been measured. Validating the XA's performance and its use in DUNE's FD--VD module, five configurations have been tested (see Table~\ref{tab:wls_configurations}). All XAs are instrumented with SiPMs from Fondazione Bruno Kessler (Triple--Trench, with pixel size \SI{50}{\micro\metre} and a total effective area of \SI{36}{\square\milli\metre}). 

\section{Instrumentation and Methodology} \label{sec:methodology} 
In the following, CIEMAT's neutrino physics group implementation of a cryogenic setup for the XA PDE measurement is described.

\subsection{Cryogenic Setup} \label{sec:ciemat_setup}
A $\sim\SI{300}{\liter}$ cylindrical vessel hosts another inner ``cassette'' vessel~($\sim\SI{80}{\liter}$) to contain the XA (see Figure~\ref{fig:xa_graphic} (right)). The larger volume is filled with liquid nitrogen~(LN$_2$) serving as a cold reservoir to liquefy GAr by thermal contact inside the inner squared vessel. Before starting the data taking, successive vacuum cycles are performed to minimize outgassing during the filling process which might worsen the LAr purity significantly. The setup is designed to maintain constant operating conditions thanks to a slow control system and automated pressure valves and pumps that perform the nitrogen filling of the outer vessel.

\subsection{Calibration Box} \label{sec:calibration_box}
\begin{figure}[t!]
    \centering
    \includegraphics[width=.35\textwidth]{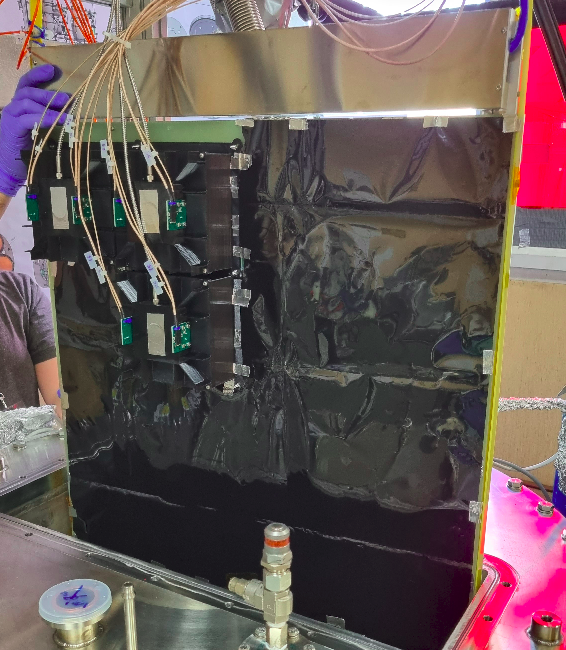}
    \quad
    \includegraphics[width=.55\textwidth]{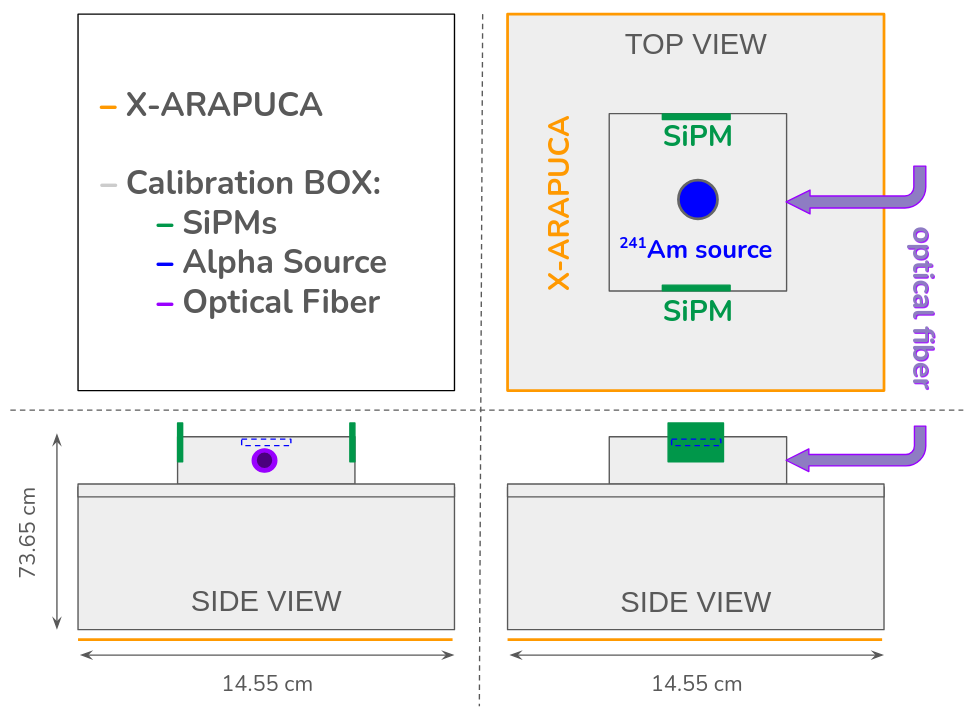}
	\caption{(left) Insertion of an XA inside the cryogenic vessel. (right) Setup scheme of the calibration box hosting the reference sensors together with the $^{241}$Am source and the inlet for the optical fiber used in calibration.\label{fig:xa_insertion}}
\end{figure}

The XA is submerged together with three calibration black boxes (see Figure~\ref{fig:xa_insertion}), placed on the 3 uniquely distinct pTP--coated--tile arrangements of the XA. Each of these boxes hosts two reference HPK VUV4 SiPMs (S13370--6075CN) which are designed for VUV detection and stable performance at cryogenic temperature, making it possible to directly detect the scintillation light produced in LAr by alpha particles emitted from a \SI{54.53(82)}{\becquerel} $^{241}$Am source~\cite{alamillo2023validation}. Measurements at CIEMAT~\cite{perez2024measurement} have provided an accurate PDE value of the VUV4 SiPMs of \num{12.69 \pm 1.12}\% for \SI{127}{\nano\meter} at \SI{87}{\kelvin} and \SI{4}{\volt} overvoltage~(OV).The emitted $\alpha$-particles deposit their energy in the LAr contained in the calibration box and produce photons that reach the XA and the reference sensors. A black sheet covers the rest of the XA substrates, ensuring that only photons produced inside a calibration box are being detected. An optical fiber guides light from an external laser source (blue light at \SI{405}{\nano\meter}) that can be operated by a waveform generator providing simultaneously a precise (\unit{\nano\second} wide) pulse of light and the appropriate trigger to the DAQ to perform the necessary calibration runs (see Section \ref{sec:calibration}).

\section{PDE Measurement}\label{sec:pde_method}
Comparing the XA measurement to the amount of light collected by the reference sensors (providing a measurement of the absolute light) is a reliable method to compute the XA's detection efficiency because it is less sensitive to changing LAr--purity conditions that can have a strong impact on the amount of light produced on each alpha--interaction. The computation is divided into several parts.

\subsection{Sensor Calibration} \label{sec:calibration}
A low--intensity laser pulse provided by the external laser and fed into the calibration box via the optical fiber is used to measure the gain, the signal--to--noise ratio~(SNR) and the cross--talk ($\rm P_{XT}$) of the XA and reference SiPMs. The calibrations show a SNR~$>2$ in all cases, which complies with DUNE's detector requirements. Finally, the cross--talk probability is also computed with the Vinogradov method~\cite{vinogradov2009probability} from calibration data. The average result from the calibration of both the XA and reference SiPMs across all runs of data--taking are shown in Table \ref{tab:sipm_calibration}.

\begin{table}[t!]
    \resizebox{.51\textwidth}{!}
    { 
        \begin{tabular}{cccc}
            \multicolumn{4}{c}{ FBK TT (XA's instrumented SiPMs)}       \\\hline
            OV (V) & Gain ($·10^{5}$) & $\rm S/N$ & $\rm P_{XT}$ (\%) \\ \hline \hline
            3.5 & $3.52\pm 0.03$  & 3.1 $\pm$ 0.8 & 12.8 $\pm$ 0.3 \\
            4.5 & $4.48\pm 0.04$  & 3.7 $\pm$ 0.9 & 18.4 $\pm$ 0.4 \\
            7.0 & $6.89\pm 0.06$  & 5.4 $\pm$ 1.3 & 32.6 $\pm$ 0.7 \\ \hline
        \end{tabular}
    }
    \resizebox{.47\textwidth}{!}
    {
        \begin{tabular}{cccc}
            \multicolumn{4}{c}{HPK VUV4 (reference SiPMs) }  \\\hline
            OV (V) & Gain ($·10^{6}$) & $\rm S/N$ & $\rm P_{XT}$ (\%) \\ \hline \hline
            3.0 & \num{6.0(2)}  & \num{6.3(1.4)} & \num{18(2)} \\
            4.0 & \num{8.0(3)}  & \num{6.5(2.1)} & \num{25(3)} \\
            5.0 & \num{10.1(5)} & \num{6.9(2.6)} & \num{33(4)} \\ \hline
        \end{tabular}    	
    }
    \caption{Average SiPM calibration results.}
    \label{tab:sipm_calibration}
\end{table}

\subsection{Charge Integration} \label{sec:charge_computation}
\begin{figure}[t!]
    \centering
    \includegraphics[width=.46\textwidth]{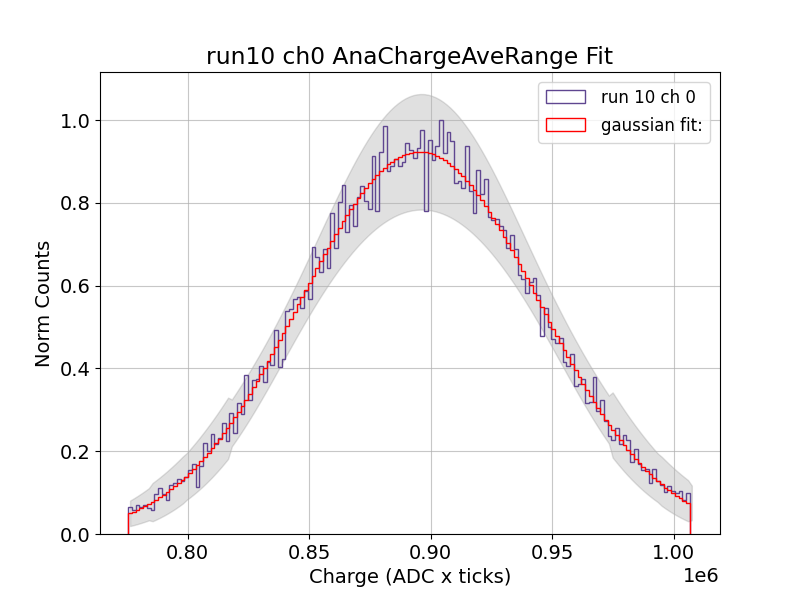}
    \includegraphics[width=.52\textwidth]{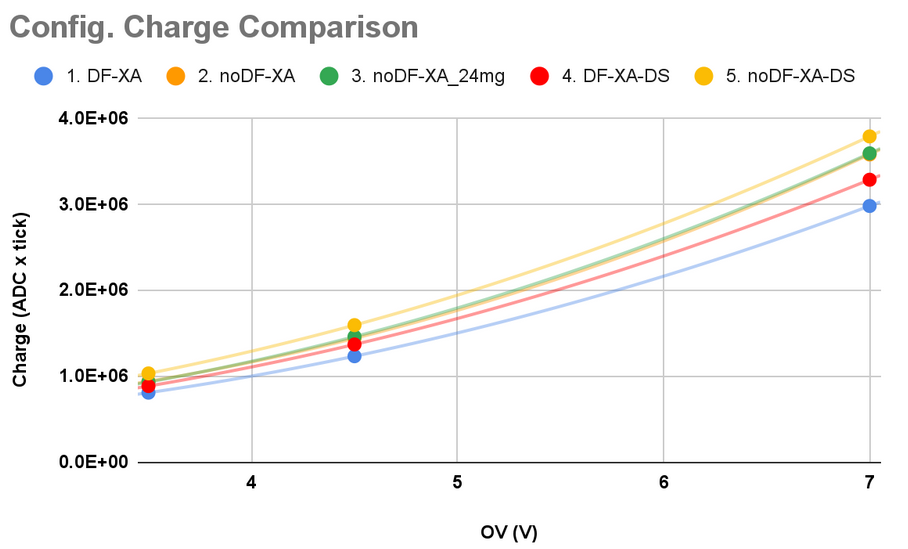}
    \caption{(left) Example gaussian--fit of integrated scintillation charge for the XA. (right) Charge comparison between XA--configurations and OV values.\label{fig:scintillation}}
\end{figure}

With the calibrated charge response of the sensors, the average number of PEs per alpha interaction can be computed by fitting the charge spectrum to a Gaussian (see Figure~\ref{fig:scintillation} (left)). A summary of the averaged XA collected charges vs over voltage can be seen in Figure~\ref{fig:scintillation} (right) for the tested configurations. In the case of the reference sensors, the individual charges are first converted to PE values (using the calibrated gain factor) and then summed on an event--by--event basis and gaussian--fitted in common to mitigate the non-gaussian response of the individual SiPMs caused by their asymmetric position with respect to the alpha source.

\subsection{PDE Computation} \label{sec:error_estimation}
The efficiency of the XA ($\epsilon_\text{XA}$) is obtained from the ratio of detected photoelectrons~(PE) in the XA wrt reference SiPMs, the cross--talk correction factors, the known efficiency of the reference SiPM’s ($\epsilon_\text{Ref. SiPM}$) and the correction for geometrical acceptance of the calibration box (evaluated to $f_\text{geo} = 0.047 \pm 0.001$), as is shown in equation~\ref{eq:mtha} (PDE values for each XA configuration are summarized in Table~\ref{tab:results}). 

\begin{equation}
    \epsilon_\text{XA} = \frac{\#\text{PE}_\text{XA}}{\#\text{PE}_\text{Ref. SiPM}}\cdot  \frac{f_\text{XT} (\text{XA})}{f_{XT} (\text{Ref. SiPM})} \cdot \epsilon_\text{Ref. SiPM}\cdot f_\text{geo}\,.
    \label{eq:mtha}
\end{equation}

\section{Results and Error Estimation} \label{sec:results}
From the results presented in Table~\ref{tab:results}, one can conclude that the PDE values average to $\sim4\%$, which is above the experiment's requirements. We observe the $\sim10 - 20\%$ increment in PDE predicted by simulation for the configurations without dichroic filter. Additionally, no significant efficiency loss is measured between the single-- and double--sided configurations, from which an efficient light confinement inside the WLS bar and a poor recovery of the exiting photons can be concluded. Finally, the measured alternative WLS bar thickness and chromophore concentration show no further improvement to the PDE.

\begin{table}[ht!]
\centering
    \begin{tabular}{clccc} 
        \multicolumn{5}{c}{VD--XA PDE Results}       \\\hline
        & Config          & OV \SI{3.5}{\volt} & OV \SI{4.5}{\volt} & OV \SI{7}{\volt}\\ \hline \hline
        1 & DF-XA         & 3.2 $\pm$ 0.2      & 3.7 $\pm$ 0.3      & 4.7 $\pm$ 0.3   \\ 
        2 & DF-XA-DS      & 3.5 $\pm$ 0.3      & 4.0 $\pm$ 0.4      & 5.0 $\pm$ 0.5   \\
        3 & noDF-XA       & 3.9 $\pm$ 0.4      & 4.5 $\pm$ 0.4      & 5.8 $\pm$ 0.6   \\
        4 & noDF-XA-DS    & 3.8 $\pm$ 0.4      & 4.5 $\pm$ 0.4      & 5.6 $\pm$ 0.6   \\
        5 & noDF-XA\_24mg & 3.6 $\pm$ 0.4      & 4.3 $\pm$ 0.4      & 5.5 $\pm$ 0.6   \\\hline
    \end{tabular}
    \caption{Results for the XA absolute efficiency $\epsilon_\text{XA}$ for different configurations and OV values. \label{tab:results}}
\end{table}

The main source of uncertanty is the reference SiPM's PDE, adding to the result with a relative $\pm 8.85\%$ ($\epsilon_\text{Ref. SiPM}$, from publication~\cite{perez2024measurement}). The final error of the measurement~$(\Delta\epsilon)$ is evaluated from the squared root of the summed squares of the uncertanties. The cross--talk correction factors~$f_\text{XT} = (1-\text{P}_\text{XT})$ and follow the derivation of the coefficient of duplication $(K_\text{dup})$ as described in~\cite{vinogradov2009probability}.


\bibliographystyle{JHEP}
\bibliography{biblio.bib}

\end{document}